\documentclass[12pt,reqno,a4paper]{amsart}
\usepackage{graphicx} % Required for inserting images
\usepackage{amssymb}
\usepackage{bbm}
\usepackage{amsmath}        % extensions for typesetting of math
\usepackage{amsfonts}       % math fonts
\usepackage{amsthm}         % theorems, definitions, etc.
\usepackage{bbding}         % various symbols (squares, asterisks, scissors, ...)
\usepackage{bm}             % boldface symbols (\bm)
\usepackage{graphicx}       % embedding of pictures
\usepackage{fancyvrb}       % improved verbatim environment
\usepackage{tikz-cd}
\usepackage{rotating}
\usepackage{dcolumn}        % improved alignment of table columns
\usepackage{booktabs}       % improved horizontal lines in tables
\usepackage{paralist}       % improved enumerate and itemize
\usepackage{xcolor}         % typesetting in color
\usepackage{bbold}
\usepackage{hyperref}
\usepackage{breqn}
\usepackage{subcaption} %
\usepackage{cancel}

\usepackage{mathtools}
\newcommand{\defeq}{\vcentcolon=}

\newtheorem{claim}{Claim}[section]
\newtheorem{theorem}[claim]{Theorem}

\newtheorem{lemma}[claim]{Lemma}
\newtheorem{corollary}[claim]{Corollary}
\newtheorem{remark}[claim]{Remark}

\newcommand*\Laplace{\mathop{}\!\mathbin\bigtriangleup}

\DeclareMathOperator*{\esssup}{ess\,sup}

\newcommand{\ee}{\mathrm{e}}
\newcommand{\D}{\mathrm{d}}

\newcommand{\N}{\mathbb{N}}
\newcommand{\R}{\mathbb{R}}

\title{Spectral properties of twisted soft waveguides}
\author[P.~Exner]{Pavel Exner}
\address{Doppler Institute for Mathematical Physics and Applied Mathematics,
Czech Technical University, B\v{r}ehov\'a 7, 11519 Prague, Czechia, and
Department of Theoretical Physics, NPI, Academy of Sciences,
25068 \v{R}e\v{z} near Prague, Czechia}
\email{exner@ujf.cas.cz}

\author[D.~Spitzkopf]{David Spitzkopf}
\address{Department of Theoretical Physics, NPI, Academy of Sciences,
25068 \v{R}e\v{z} near Prague, Czechia}
\email{spitzkopf@ujf.cas.cz}
\thanks{Corresponding author: D.~Spitzkopf (spitzkopf@ujf.cas.cz)}

\begin{document}

\begin{abstract}
We discuss a model of a soft waveguide described by Schr\"odinger operator whose
potential is a twisted straight channel of non-circular profile. We show that a
compactly supported twist makes the operator subcritical, the property being
expressed in terms of a Hardy-type inequality. On the other hand, a periodic
twist raises the essential spectrum and its local slowdown may give rise to
isolated eigenvalues below the essential spectrum threshold.
\end{abstract}
\maketitle

%%%%%%%%%%%%%%%%%%%%%%
\section{Introduction}
%\label{sec:Intro}

Among relations between spectral properties and the geometry, the dynamics in the vicinity of curves plays an important role. It is well known, for instance, that the Dirichlet Laplacian in a tubular region which is bent but asymptotically straight has nonempty discrete spectrum due to the effective attractive interaction coming from the tube geometry; such operators are used as models of quantum waveguides cf.~\cite{EK15} for a survey. Other geometric perturbations can have different consequences. In particular, twisting of tube of a noncircular profile gives rise to an effective interaction which is of a \emph{repulsive} nature. This effect was first observed in \cite{CB96} and put on a solid mathematical basis in \cite{EKK08}.

Recently another waveguide model, labeled as `soft' and more realistic from the
physics point of view, has attracted attention. In this case the confinement is
not realized by a hard wall modeled by Dirichlet boundary conditions, but rather by an attractive potential in the form of a channel. The effect of `binding by bending' appeared to be robust; its existence was demonstrated in this setting too, both in two \cite{Ex20, KKK21, EV24} and three \cite{Ex22} dimensions. The question naturally arises whether a twisting of a three-dimensional waveguide modeled by a potential channel of an asymmetric profile produces an effective repulsion. The aim of this paper is to provide an affirmative response to this question.

%%%%%%%%%%%%%%%%%%%%%%
\section{Problem setting and main results}
\label{sec:Main}
\setcounter{equation}{0}

By a straight soft quantum waveguide in $\R^3$ we mean a system described by
Schr\"odinger operator with a translationally invariant potential of the profile
$W:\R^2 \to \R$; its \emph{twisting} means that the profile rotates as it moves
along the axis. To express it in formal terms, we consider the operator $H
\defeq - \Laplace + V$ with the form domain $H^1(\R^3)$, and the potential $V:\,
V(s,t) \defeq W(A^{-1}(s)t)$, where $(s,t) \in \R\times \R^2$ and $A(s)\in
SO(2)$ is the rotation by an angle $\theta(s)$ with a reasonably regular
dependence of the longitudinal coordinate, say, $\theta\in W^{2,\infty}_{\mathrm{loc}}(\R)$. The operator $H$ is uniquely associated with the quadratic form $\mathbb{q}[\cdot]$ defined by
% --------------
\begin{equation}
\label{eq:quadrForm}
\mathbb{q}[\psi] \defeq \int_{\R^3} |\nabla \psi(s,t) |^2 \D s \D t +
\int_{\R^3} W(A^{-1}(s)t) |\psi(s,t)|^2 \D s \D t.
\end{equation}
% -------------- %
We are interested primarily in \emph{local} profile potentials in the sense that
$W(t)\to 0$ as $|t| \to \infty$, however, some results extend to other potentials used to model quantum waveguides, cf.~Remark~\ref{rem:divprofile} below. The most important assumption we impose is that
% -------------- %
\begin{itemize}
\item[$\mathbf{(A)}$] $W$ is \emph{not} rotationally symmetric and $H_t \defeq -
    \Laplace_t + W$ has a simple isolated ground state eigenvalue associated with the eigenfunction $\phi_0 >0$.
\end{itemize}
% -------------- %
In addition, we will make some not very restrictive assumptions about the local
integrability of $W$ to ensure the self-adjointness of operators $H$ and $H_t$ by means of the KLMN theorem; these are specified in Section~\ref{sec:Basics} below and the elaborated hypothesis is marked as $\mathbf{(A_{\mathrm{sa}})}$. To state the results we denote the energy of the ground state of $H_t$ mentioned above as $\epsilon_0$. We also introduce the symbol $\eta \defeq \inf(\sigma(H_t)\setminus \lbrace\epsilon_0\rbrace) - \epsilon_0$ for the spectral gap of $H_t$; by assumption we have $\eta > 0$.

Excluding the trivial situation when the twist is zero, we focus on two classes, namely local ones and local perturbation of a periodic twist, specified in the following way:
% -------------- %
\begin{itemize}
\item[$\mathbf{(T1)}$] $\theta'$ is compactly supported and there exists a
    nonempty compact interval $J_0\subset \mbox{supp }\theta'$ such that $\min_{s\in J_0}
    |\theta'(s)| \geq \theta_{\mathrm{min}} > 0$,
\item[$\mathbf{(T2)}$] $\theta'_p = \beta_0 + \delta$ with $\beta_0 \neq 0$
    and $\delta$ compactly supported.
\end{itemize}
% -------------- %
Let us begin with the main results for the operators satisfying $\mathbf{(T1)}$, a Hardy-type inequality showing that a nonzero local twist makes the operator $H$ subcritical. We can do it in two ways, first as a lower bound with a two-variable weight:
% -------------- %
\begin{theorem}
\label{theo:TwoParamHardy}
Assume $\mathbf{(A_{\mathrm{sa}})}$ and $\mathbf{(T1)}$, then there are numbers \mbox{$\tilde{c}_{\mathrm{H}}>0$} and $s_0 \in J_0$ such that for any $\psi \in H^1(\R^3)$ we have
% -------------- %
$$
\mathbb{q}[\psi] - \epsilon_0 \|\psi\|^2 \geq \tilde{c}_{\mathrm{H}} \int_{\R\times\R^2}
\frac{|\psi(s,t)|^2}{(1+(s-s_0)^2)(1+|t|^2)}\, \D s \D t.
$$
% -------------- %
\end{theorem}
% -------------- %
Furthermore, modifying the constant, one can get a stronger result, an inequality having a weight without the transversal decay:
% -------------- %
\begin{theorem}
\label{theo:MainHardy}
Under the same assumptions, put $c_{\mathrm{H}}\defeq \frac12 \Gamma(\tilde{c}_{\mathrm{H}})$, where $\Gamma(\nu) \defeq \min \big\lbrace \frac{\nu m_0}{4}, \frac{\eta
m_0}{2(4+m_0)} \big\rbrace$ with $m_0 \defeq \int \phi_0^2 (1+|t|^2)^{-1}\D t$; then we have
% -------------- %
$$
\mathbb{q}[\psi] - \epsilon_0 \|\psi\|^2 \geq c_{\mathrm{H}} \int_{\R \times \R^2}
\frac{|\psi(s,t)|^2}{1+(s-s_0)^2}\, \D s \D t \quad\text{for any}\quad \psi \in H^1(\R^3)
$$
% -------------- %
\end{theorem}
% -------------- %

As we will show in Sections \ref{sec:ProofTwoPar} and \ref{sec:ProofMixed} below, the transversal decay in the first theorem allows us to control the angular derivative, the stronger bound in Theorem~\ref{theo:MainHardy} can be obtained by means of the coercivity argument,
% -------------- %
\begin{equation}
\label{eq:LowerBoundViaHt}
\mathbb{q}[\psi] - \epsilon_0 \|\psi\|^2 \geq \int_\R w_s\,(\psi(s,\cdot),
(H_t - \epsilon_0) \psi(s,\cdot))_{L^2(\R^2)} \D s \quad\forall \psi\in H^1(\R^3),
\end{equation}
% -------------- %
with a weight $w_s\in(0,1]$. The intermediate result from Theorem~\ref{theo:TwoParamHardy} is needed at that, because the right-hand side of \eqref{eq:LowerBoundViaHt} vanishes, e.g., for functions of the form $\psi(s,t) = u(s)\phi_0(t)$ with $u \in C^\infty_0(\R)$.

We can also combine the two results in the following way:
% -------------- %
\begin{corollary}
\label{cor:WeightCombined}
Under the assumptions of the above theorems, we have
% -------------- %
$$ %\begin{equation}
%\label{eq:genHardyweight}
\mathbb{q}[\psi] - \epsilon_0 \|\psi\|^2 \geq \int_{\R\times\R^2}
\rho(s,t) |\psi(s,t)|^2\, \D s \D t  \quad \forall \psi \in H^1(\R^3),
$$ %\end{equation}
% -------------- %
where
% -------------- %
$$
    \rho(s,t) := \frac12 (1+(s-s_0)^2)^{-1} \max \big \lbrace \tilde{c}_{\mathrm{H}} (1+|t|^2)^{-1},
    c_{\mathrm{H}}\big \rbrace.
$$
% -------------- %
\end{corollary}
% -------------- %

The second main result is complementary to the first one: a perturbation of a uniformly twisted guide by slowing locally the twist rate \emph{in the mean gives} rise to an effective attraction which produces a nonempty discrete spectrum below the (shifted) essential spectrum threshold. We get a counterpart to the analogous results for hard-wall waveguides, cf.~\cite{EKK08} and \cite{EK05}, respectively, where the Hardy inequality and the binding by a local twist slowdown are two faces of one repulsion mechanism; in the soft setting the anisotropy of $W$ plays the role which the tube cross section anisotropy played there.

We thus adopt hypothesis $\mathbf{(T2)}$ and use the symbol $H_{\beta_0}$ for the corresponding Hamiltonian of the guide with a constant twist $\theta'=\beta_0$; by $H_{\beta}$ we denote the one with the varying twist rate $\theta'_p$. Furthermore, we will require that
% -------------- %
\begin{itemize}
\item[$\mathbf{(B)}$] the operator $h_{\beta_0} \defeq -\Laplace_t + W
	+\beta_0^2 p_\varphi^2$ has an isolated ground state $\phi_{\beta_0}$ with energy $\epsilon(\beta_0)$.
\end{itemize}
% -------------- %
\begin{remark}
\label{rem:BHappensList}
 Since $h_{\beta_0} \geq H_t$ holds in the form sense, $\inf
	\sigma_{\mathrm{ess}}(h_{\beta_0}) \geq \inf \sigma_{\mathrm{ess}}(H_t)
	= 0$. The validity of assumption $\mathbf{(B)}$ is thus guaranteed if
$\inf \sigma(h_{\beta_0}) < 0$, the infimum being then a discrete eigenvalue. This holds in two particular situations, one representing a restriction to the value of $\beta_0$, the other a restriction on the potential $W$:
% -------------- %
\begin{enumerate}[(i)]
\item for any $W$ satisfying $\mathbf{(A_{\mathrm{sa}})}$, provided that $\beta_0^2\|p_\varphi \phi_0\|^2 < |\epsilon_0|\|\phi_0\|^2$,
\item for every $\beta_0$, provided that $(u,H_tu) < 0$ holds for some radial $u \in H^1(\R^2)$, which is the case, in particular, if $W_+ \in L^1(\R^2)$ and $\int_{\R^2} W(t)\, \D t < 0$,
\end{enumerate}
% -------------- %
as we will demonstrate in Remark~\ref{rem:BHappensDetail} below.

\end{remark}

% -------------- %
\begin{theorem}
\label{theo:SlowDownMean}
Assume $\mathbf{(A_{\mathrm{sa}})}$, $\mathbf{(B)}$ and $\mathbf{(T2)}$, then $\sigma_{\mathrm{disc}}\left(H_\beta \right) \neq \emptyset$ holds provided that
% -------------- %
$$
\int_\R \left(\theta'_p(s)^2 - \beta_0^2 \right) \D s < 0.
$$
% -------------- %
\end{theorem}
% -------------- %

The article is organized as follows. In Section~\ref{sec:Basics} we review the self-adjointness, basic spectral properties and the passage to a co-rotating frame. In Section~\ref{sec:ProofTwoPar} we present the proof of Theorem~\ref{theo:TwoParamHardy}, except for Lemma~\ref{lemma:mixed}, which is relegated to Section~\ref{sec:ProofMixed}. After that, in Section~\ref{sec:ProofThm2}, we prove Theorem~\ref{theo:MainHardy}, and finally, Theorem \ref{theo:SlowDownMean} will be proven in Section~\ref{sec:PoTSlowDown}.

%%%%%%%%%%%%%%%%%%%%%%%%%%
\section{Basic Properties}
\label{sec:Basics}
\setcounter{equation}{0}

In this section we review basic properties of the considered systems and prove some useful lemmata. First we make more precise the profile potential specification:
\begin{itemize}
\item[$\mathbf{(A_{\mathrm{sa}})}$] In addition to the assumptions made above, $\mathbf{(A)}$ and $W(t)\to 0$ as $|t|\to\infty$, let us suppose that $W \in L^p_{\mathrm{loc}}(\R^2)$ for some $p>1$.
\end{itemize}
% -------------- %
In view of the vanishing limit, there is an $R_0 >0$ such that $|W(t)| \leq 1$ holds outside the ball $B_{R_0}$ centered at the origin, hence for every $x \in \R^2$ and $r \in (0,1]$ we have
% -------------- %
\begin{align*}
\int_{|x-y|<r} \ln \frac{1}{|x-y|}\, |W(y)|\, \D y \leq \:&
\|W\|_{L^p(B_{R_0+1})} \|\, \ln (1/|\cdot|) \|_{L^{p'}(B_r)} \\
& +\pi r^2 \left(\textstyle{\ln\frac1r + \frac12} \right) \xrightarrow[r\to0]{} 0
\end{align*}
% -------------- %
by H\"older inequality, uniformly in $x$, where $p' < \infty$ since $p>1$. Thus $W$ belongs to the Kato class $\mathcal{K}_2$, cf.~\cite[(A13)]{Si82}, and consequently, $|W|$ is infinitesimally form-bounded with respect to $-\Laplace_t$; in particular, $H_t = -\Laplace_t + W$ is self-adjoint and semibounded with form domain
$\mathcal{Q}(H_t) = H^1(\R^2)$ by \cite[Thm A.2.7]{Si82}.

Since the twist $A(s) \in SO(2)$ is an isometry for each s, the substitution $t \mapsto A(s)t$ leaves both sides of the fiber-wise bound invariant, in other words, $\int_{\R^2} |V(s,t)| |\psi(s,t)|^2 \D t \leq \epsilon \|\nabla_t \psi(s,\cdot)\|^2_2 + C_\epsilon \|\psi(s,\cdot)\|^2_2$ for almost every $s$, with $\epsilon$ and $C_\epsilon$ independent of $s$. Integrating over $s$ and using the inequality $\|\nabla_t \psi \|_2 \leq \| \nabla \psi \|_2$, we
conclude that $|V|$ is form-bounded with an arbitrarily small bound relative to $\|\nabla \psi\|^2_2$. By the KLMN theorem \cite[Thm X.17]{RS}, the form $\mathbb{q}$ is therefore closed and bounded from below on $H^1(\R^3)$ and it defines a unique self-adjoint and semi-bounded operator $H$.

We will make use of a `straightening' transformation defined by the unitary $U$,
acting as $(U\psi)(s,t) \defeq \psi (s,A^{-1}(s)t)$, in other words, as rotation
in the transverse plane by the angle $\theta(s)$ for each $s$. The quadratic form of interest is given by \eqref{eq:quadrForm}; looking at the longitudinal derivative  $\partial_s$ first, we have $A(s)= {\scriptsize \begin{pmatrix} \cos\theta(s) & -\sin\theta(s) \\ \sin\theta(s) & \cos\theta(s) \end{pmatrix}}$ and
% ------------- %
\begin{align*}
\partial_s(\psi(s,A^{-1}(s)t)) = & \: (\partial_s\psi)(s,A^{-1}(s)t) \\ & - \nabla_t\psi(s,A^{-1}(s)t)\cdot(A^{-1}(s)A'(s)A^{-1}(s)t),
\end{align*}
% ------------- %
where
% ------------- %
$$
A'(s) = \theta'(s)\begin{pmatrix}0 & -1 \\ 1 & 0\end{pmatrix}A(s).
$$
% ------------- %
Substituting this and using the angular derivative $\partial_\varphi := t_1\partial_{t_2}-t_2\partial_{t_1}$, we can write the above expression concisely as
% ------------- %
$$
\partial_s(\psi(s,A^{-1}(s)t))
=
((\partial_s-\theta'(s)\partial_\varphi)\psi)(s,A^{-1}(s)t).
$$
% ------------- %
Note that the correspondence between $\nabla_t$ and $A(s) \nabla_t$ is pointwise norm-preserving and both the  $(1+(s-s_0)^2)^{-1}$ and $w_t:=(1+|t|^2)^{-1}$ are $U$-invariant. The main purpose of the transformation is to use the relation $(\psi, U^{-1}VU\psi) = (\psi, W\psi)$ to move the angular dependence from the potential term to the kinetic one, rewriting \eqref{eq:quadrForm} as
% ------------- %
\begin{equation} \label{straightform}
\mathbb{q}[\psi ] = \int_{\R \times \R^2} \left(|(\partial_s - \theta'(s) \partial_\varphi) \psi|^2 + |\nabla_t \psi|^2 +  W(t) |\psi|^2\right)(s,t)\, \D s \D t,
\end{equation}
% ------------- %
which is the form we are going to use from now on.

Next we introduce a quantity related to the spectral gap which will be needed in the following; it is defined as
% ------------- %
$$ %\begin{equation}
%\label{define:lambda}
\lambda \defeq \inf_{\psi \in \mathcal{Q}(H_t) \setminus \lbrace 0 \rbrace}
\frac{(\psi, (H_t - \epsilon_0) \psi) + \| \partial_\varphi \psi
\|^2_{L^2(\R^2,w_t\D t)}}{\|\psi \|^2_{L^2(\R^2,w_t\D t)}}
$$ %\end{equation}
% ------------- %
with the weight $w_t$ in the fiber space indicated above. The importance of the second term in the numerator comes from the following simple observation:
% ------------- %
\begin{lemma}
\label{lemma:PhiNoRotSymm}
$\partial_\varphi \phi_0 = 0$ holds if and only if $W(\cdot)$ is rotationally symmetric.
\end{lemma}
% ------------- %
\begin{proof}
If $W$ has rotational symmetry we can separate variables in polar coordinates, and since the ground state belongs to the zero angular momentum subspace, the solution $\phi_0$ is radial. On the other hand, should $\phi_0$ be radial and $W$ not (a.e., as an element of $L^2(\R^2)$), then the relation $W= \frac{(\Laplace_t+\epsilon_0)\phi_0}{\phi_0}$ would lead to a contradiction.
\end{proof}
% ------------- %
\begin{lemma}
\label{lemma:lambdaPositive}
Suppose that $\partial_\varphi \phi_0 \neq 0$ and $\eta > 0$, then $\lambda > 0$.
\end{lemma}
% ------------- %
\begin{proof}
By \emph{reductio ad absurdum} assuming that $\lambda = 0$. In that case there exists a sequence $\{\phi_n\}_{n\in\N} \subset \mathcal{Q}(H_t) = H^1(\R^2)$ of unit norm vectors, $\|\phi_n\|^2_{L^2(\R^2,w_t\D t)} = 1$, such that
% ------------- %
$$
(\phi_n, (H_t - \epsilon_0)\phi_n)
+ \|\partial_\varphi \phi_n\|^2_{L^2(\R^2,w_t \D t)} \longrightarrow 0
$$
% ------------- %
as $n\to\infty$; both terms are obviously nonnegative, so each must tend to zero separately. Decompose $\phi_n = a_n \phi_0 + \psi_n$ with $a_n \defeq (\phi_0, \phi_n)$ and $(\psi_n, \phi_0) = 0$. Since $\phi_0$ is the ground state and $\epsilon_0$ is simple with the gap $\eta > 0$, we have
% ------------- %
$$
(\phi_n, (H_t - \epsilon_0)\phi_n)
= (\psi_n, (H_t - \epsilon_0)\psi_n)
\geq \eta \, \|\psi_n\|^2_{L^2},
$$
% ------------- %
thus $\|\psi_n\|_{L^2} \to 0$. Moreover, combining the infinitesimal form bound $\int W_-|\psi_n|^2\D t \leq \epsilon \|\nabla_t \psi_n\|^2 + C_\epsilon \|\psi_n\|^2_{L^2}$ valid for all $\epsilon \in (0,1)$ with the fact that $(\psi_n, (H_t - \epsilon_0)\psi_n) \to 0$ and $\|\psi_n\|_{L^2} \to 0$, we get $\|\nabla_t \psi_n\| \to 0$; hence $\psi_n \to 0$ holds in $H^1(\R^2)$ too.

Next, from $\|\phi_n - a_n\phi_0\|_{L^2(\R^2,w_t\D t)} = \|\psi_n\|_{L^2(\R^2,w_t\D t)} \leq \|\psi_n\|_{L^2} \to 0$ and the assumed normalization we infer that
% ------------- %
$$
\big|\, 1 - |a_n|\sqrt{m_0} \,\big|
= \big|\, \|\phi_n\|_{L^2(\R^2,w_t\D t)} - \|a_n\phi_0\|_{L^2(\R^2,w_t\D t)} \,\big|
\leq \|\psi_n\|_{L^2(\R^2,w_t \D t)}
$$
% ------------- %
with $m_0 := \|\phi_0\|^2_{L^2(\R^2,w_t\D t)} > 0$ tends to zero. Thus $|a_n|^2 m_0 \to 1$ which means, in particular, that  $\{a_n\}$ is bounded and there is a subsequence along which $a_n \to a$ with $|a| = m_0^{-1/2} \neq 0$, and this in turn implies $\phi_n = a_n\phi_0 + \psi_n \to a\phi_0$ in $H^1(\R^2)$. For any $u \in H^1(\R^2)$ we have
% ------------- %
$$
\|\partial_\varphi u\|_{L^2(\R^2,w_t\D t)}
\leq \big\| |t|\, w_t^{1/2} \nabla_t u \big\|_{L^2}
\leq \|\nabla_t u\|_{L^2},
$$
% ------------- %
since $|t|^2 w_t \leq 1$, hence $\partial_\varphi : H^1(\R^2) \to L^2(\R^2,w_t\D t)$ is bounded. Combining this with the convergence $\phi_n \to a\phi_0$ in $H^1$, we get
% ------------- %
$$
\|\partial_\varphi \phi_n\|_{L^2(\R^2,w_t\D t)}
\longrightarrow |a| \, \|\partial_\varphi \phi_0\|_{L^2(\R^2,w_t\D t)}.
$$
% ------------- %
We know that the left-hand side tends to zero, and $a \neq 0$, so $\partial_\varphi \phi_0 = 0$ must hold in $L^2(\R^2,w_t\D t)$, and since $w_t > 0$, this means that \mbox{$\partial_\varphi \phi_0 = 0$}, which is the sought contradiction.
\end{proof}
% ------------- %
\begin{remark}
{\rm
The proof given above does not provide the value of $\lambda$. An explicit lower
bound to it can be obtained in terms of the Friedrichs angle between
$\mbox{span } \lbrace \phi_0 \rbrace$ and the radial subspace of $L^2(w_t \D
t)$,  cf. \cite{Ha69}, but we will not pursue this idea.
}
\end{remark}
% ------------- %

In what follows we employ a ground-state substitution $\psi=\phi_0 u$, requiring $u \defeq \psi/\phi_0$ to be admissible, meaning that $\phi_0^{-1} \in L^\infty_{\rm{loc}}(\R^2)$. Furthermore, the shifted form must have a fiber decomposed lower bound. This is ensured by the following lemma.
% ------------- %
\begin{lemma}
\label{lemma:GroundStateFiber}
Under the assumption $\mathbf{(A_{\mathrm{sa}})}$, the following holds:
% ------------- %
\begin{itemize}
    \item[(a)] $\phi_0^{-1} \in L^\infty_{\rm{loc}}(\R^2)$.
    \item[(b)] For every $\psi \in Q(\mathbb{q})$ and every measurable weight $\omega : \R \to [0,1]$ one has
% ------------- %
$$ %\begin{equation}
\mathbb{q}[\psi] - \epsilon_0\|\psi\|^2
\geq \int_\R \omega(s)\, h[\psi(s,\cdot)]\, \D s,
$$ %\end{equation}
% ------------- %
where $h[g] \defeq (g,(H_t-\epsilon_0)g)
= \|\nabla_t g\|^2_{L^2(\R^2)} + \int_{\R^2}(W-\epsilon_0)|g|^2\,\D t$
for $g \in H^1(\R^2) = \mathcal{Q}(H_t)$.
% ------------- %
\end{itemize}
\end{lemma}
% ------------- %
\begin{proof}
As for (a), it suffices to show that $\inf_{t\in M}\phi_0(t) > 0$ holds on any compact $M \subset \R^2$. Pick $R$ such that $M\subset B_R$, the ball of radius $R$ centered at the origin. It follows from \cite[Cor. 5.3]{Tr73}, with $\Omega' = B_R$ and $\Omega= B_{R+1}$ that\footnote{In the notation of \cite{Tr73}, $\lambda = \Lambda \equiv 1$ and $g = |W- \epsilon_0| \in L_s(B_{R+1})$ with $s=p$ so that \cite[eq.~(5.1)]{Tr73} holds for any finite $t > (1-1/p)^{-1}$.}
% ------------- %
$$
\inf_{B_R} \phi_0 \geq C^{-1} \sup_{B_R}\phi_0
$$
% ------------- %
holds with a positive constant $C$ dependent on $R$. This implies
% ------------- %
$$
\inf_M \phi_0 \geq \inf_{B_R} \phi_0 \geq C^{-1} \sup_{B_R} \phi_0 > 0,
$$
% ------------- %
because $\phi_0$ is positive.

Passing to claim (b), since $|W|$ is form-bounded with respect to $-\Laplace$, the function $(s,t)\mapsto W(t)|\psi(s,t)|^2$ is integrable on $\R^3$, and at the same time $\int_\R\|\psi(s,\cdot)\|^2_{H^1(\R^2)}\,\D s \leq \|\psi\|^2_{H^1(\R^3)}$. By Fubini theorem, $\psi(s,\cdot)\in H^1(\R^2) = Q(H_t)$ for almost all $s$ and $s\mapsto h[\psi(s,\cdot)]$ is measurable and finite a.e.; moreover, $h[\psi(s,\cdot)]\geq 0$ holds because $\epsilon_0 = \inf\sigma(H_t)$. Neglecting the term $\|(\partial_s - \theta'\partial_\varphi)\psi\|^2 \geq 0$, we get
% ------------- %
\begin{align*}
\mathbb{q}[\psi] - \epsilon_0\|\psi\|^2
&\geq \int_{\R^3}|\nabla_t\psi|^2 + \int_{\R^3}(W-\epsilon_0)|\psi|^2
= \int_\R h[\psi(s,\cdot)]\,\D s \\ \nonumber
&\geq \int_\R \omega(s)\,h[\psi(s,\cdot)]\,\D s,
\end{align*}
% ------------- %
which concludes the proof.
\end{proof}

% ------------- %
\begin{remark}
\label{rem:T2localbounded}
\rm{Let $\phi_{\beta_0}$ be the ground state of $h_{\beta_0}$ from assumption $\mathbf{(B)}$. In what follows, specifically in the proof of Lemma~\ref{lemma:T2IdentifySpectra}, we will need the fact that $\phi_{\beta_0}^{-1} \in L^\infty_{\mathrm{loc}}(\R^2)$. Set $\tilde{h}_{\beta_0} \defeq h_{\beta_0} - \epsilon(\beta_0)$, which is a positive operator by the definition of $\epsilon(\beta_0)$.

By the first Beurling-Deny criterion \cite[Thm. ~XIII.50]{RS}, the bound $(|u|,\tilde{h}_{\beta_0}|u|)\leq (u,\tilde{h}_{\beta_0}u)$ holds for all $u \in L^2(\R^2)$ if and only if $\ee^{-t(h_{\beta_0}-\epsilon(\beta_0))}$ is reality preserving and the inequality
% ------------- %
$$
(u_+, \tilde{h}_{\beta_0} u_+) + (u_-, \tilde{h}_{\beta_0} u_-) \leq (u,\tilde{h}_{\beta_0}u)
$$
% ------------- %
is valid for all real-valued $u \in L^2(\R^2)$. To begin with, we rewrite $h_{\beta_0}$ as
% ------------- %
$$
h_{\beta_0} = H_t + \beta_0^2 p_\varphi^2 = -\nabla_t \cdot M\nabla_t + W,
$$
% ------------- %
where $M \defeq \mathbb{1}_{2x2} + \beta_0^2 \tau \tau^T $, with $\tau \defeq
(-t_2,t_1)$; from this form we see that both $M$ and $W$ are real. In addition,
$\mathcal{Q}(h_{\beta_0})$ is invariant with respect to complex
conjugation, thus $\ee^{-t(h_{\beta_0} - \epsilon(\beta_0))}$ is reality preserving.
We write $u=u_+ - u_-$, so that
% ------------- %
$$
(u,\tilde{h}_{\beta_0}u) = (u_+,\tilde{h}_{\beta_0}u_+) + (u_-,\tilde{h}_{\beta_0}u_-) - (u_+,\tilde{h}_{\beta_0} u_-) - (u_-,\tilde{h}_{\beta_0} u_+)\,;
$$
% ------------- %
the mixed terms vanish, because both $u_{\pm}$ and their derivatives have disjoint supports. Indeed, $\nabla_t u_+ = \mathbb{1}_{u>0} \nabla_t u$ and $\nabla_t u_- = -\mathbb{1}_{u<0} \nabla_t u$, hence we have
% ------------- %
$$
(u_+,\tilde{h}_{\beta_0} u_-) = \int_{\R^2} \left(\nabla_t u_+ \cdot M \nabla_t u_- + (W- \epsilon(\beta_0)) u_+u_- \right) \D t = 0,
$$
% ------------- %
where the second term is zero as $\mathrm{supp}\, u_+ \cap \mathrm{supp}\, u_-$
has zero measure; for $(u_-,\tilde{h}_{\beta_0}u_+)$ we simply swap $u_+$ and $u_-$, obtaining
% ------------- %
$$
(u,\tilde{h}_{\beta_0}u) = (u_+,\tilde{h}_{\beta_0}u_+) + (u_-,\tilde{h}_{\beta_0} u_-).
$$
% ------------- %
Since $|u|=u_+ + u_-$ and $\| |u|\| = \|u\|$, we have
% ------------- %
$$
(|u|,\tilde{h}_{\beta_0}|u|) = (u,\tilde{h}_{\beta_0}u) \implies   (|u|,h_{\beta_0}|u|) = (u,h_{\beta_0}u).
$$
% ------------- %
Hence, if $\phi_{\beta_0}$ minimizes the Rayleigh quotient of $h_{\beta_0}$, so does $|\phi_{\beta_0}|$, and consequently, the ground state may be chosen non-negative.

Local boundedness of the inverse follows by the same argument as for
$\phi_0^{-1}$ taking the parameters, in the notation of \cite{Tr73},
$\lambda=1$, $\Lambda = 1+\beta_0^2(R+1)^2$, $g=|W-\epsilon(\beta_0)| \in L_s(B_{R+1})$ with $s=p$ instead.
}
\end{remark}
% ------------- %

Lemma \ref{lemma:GroundStateFiber} has the following consequence.
% ------------- %
\begin{corollary}
Under $\mathbf{(A_{\mathrm{sa}})}$ and $\mathbf{(T1)}$, $\inf \sigma_{\mathrm{ess}}(H) =
\inf \sigma(H) = \epsilon_0$.
\end{corollary}
% ------------- %
\begin{proof}
Claim (b) of the above lemma with $\omega=1$ gives $\mathbb{q}[\psi] \geq
\epsilon_0 \|\psi \|^2$, hence $\inf \sigma(H) \geq \epsilon_0$. By assumption
$\mathbf{(T1)}$ the twist angle is constant, $\theta\equiv \theta_+ =
\theta(b)$, on the half-line $(b,\infty)$ for $b \defeq \sup \mbox{supp
}\theta'$. Consequently, $V(s,t) = W(A(\theta_+)^{-1}t)$ is $s$-independent there. For an arbitrary $k\in \R$ we take the standard Weyl sequence of the straight guide, $\psi_n(s,t) = n^{-1/2} \chi\left(\frac{s-a_n}{n}\right) \ee^{iks} \phi_0 (A(\theta_+)^{-1} t)$ with an appropriate mollifier $\chi \in C^\infty_0(\R)$ and $a_n \to \infty$ fast	enough so that the functions $\chi\left(\frac{\cdot-a_n}{n}\right)$ have disjoint supports in $(b,\infty)$. Since $H$ acts on those $\psi_n$ as a product operator, we have $\|(H-\epsilon_0 - k^2) \psi_n\| \to 0$ and $\psi_n \rightharpoonup0$, and therefore $\epsilon_0 + k^2 \in \sigma_{\mathrm{ess}}(H)$.
\end{proof}

We also need the analogous claim for $\Sigma(\beta_0) \defeq \inf \sigma(H_{\beta_0})$.
% ------------- %
\begin{lemma}
\label{lemma:T2IdentifySpectra}
Under $(\mathbf{A_{\mathrm{sa}}})$ and $\mathbf{(B)}$, we have $\Sigma(\beta_0) = \epsilon(\beta_0)$.
\end{lemma}
% ------------- %
\begin{proof}
We mimic the argument of \cite[Thm. ~1]{EK05}. By assumption $\mathbf{(B)}$ and Remark~\ref{rem:T2localbounded}, we can factorize any $\psi \in C^\infty_0(\R^3)$ as $\psi(s,t) = \phi_{\beta_0}(t) u(s,t)$. Writing out the quadratic form, we get
% ------------- %
\begin{align}
\label{eq:BetaZeroQuadrForm}
& \mathbb{q}_{\beta_0}[\phi_{\beta_0}u] - \epsilon(\beta_0) \|\phi_{\beta_0} u\|^2 \\ & \qquad = \int_{\R \times \R^2}\left(|K_{\beta_0}\psi|^2 + |\nabla_t \psi|^2 + W(t) |\psi|^2 - \epsilon(\beta_0) |\psi|^2 \right)\, \D s \D t \nonumber
\end{align}
% ------------- %
with $K_{\beta_0} \defeq \partial_s - \beta_0 \partial_\varphi$. We have $K_{\beta_0} \psi = \phi_{\beta_0} K_{\beta_0} u - \beta_0 u \partial_\varphi \phi_{\beta_0}$, and therefore
% ------------- %
$$
|K_{\beta_0} \psi|^2 = \phi_{\beta_0}^2 |K_{\beta_0} u|^2 + \beta_0^2 |u|^2 (\partial_\varphi \phi_{\beta_0})^2 - \beta_0 \phi_{\beta_0} \partial_\varphi \phi_{\beta_0} K_{\beta_0}|u|^2.
$$
% ------------- %
Integrating the last term on the right-hand side over $\R^2$ by parts and using $K_{\beta_0}(\phi_{\beta_0}\partial_\varphi \phi_{\beta_0}) = - \beta_0 \partial_\varphi(\phi_{\beta_0}\partial_\varphi \phi_{\beta_0})$, we get
% ------------- %
\begin{align*}
& -\beta_0 \int_{\R \times \R^2} \phi_{\beta_0} \partial_\varphi \phi_{\beta_0} K_{\beta_0} |u|^2 \D s \D t \\ & \qquad = -\beta^2_0 \int_{\R\times\R^2} |u|^2 \left[ (\partial_\varphi \phi_{\beta_0})^2 + \phi_{\beta_0} \partial_\varphi^2 \phi_{\beta_0} \right] \D s \D t,
\end{align*}
% ------------- %
which yields
% ------------- %
$$
\|K_{\beta_0}\psi \|^2 =\int_{\R\times\R^2} \phi_{\beta_0}^2 |K_{\beta_0}u|^2 \D s \D t- \beta_0^2 \int_{\R\times\R^2} |u|^2 \phi_{\beta_0} \partial_\varphi^2 \phi_{\beta_0} \D s \D t.
$$
% ------------- %
Substituting into \eqref{eq:BetaZeroQuadrForm}, we obtain
% ------------- %
\begin{align}
\label{eq:BetaZeroQuadrForm2}
\mathbb{q}_{\beta_0}[\phi_{\beta_0}u] - & \,\epsilon(\beta_0) \|\phi_{\beta_0}u\|^2 = \int_{\R\times\R^2} \big( \phi_{\beta_0}^2 |K_{\beta_0}u|^2 - \beta_0^2 |u|^2 \phi_{\beta_0} \partial_\varphi^2 \phi_{\beta_0} \nonumber  \\
& \qquad\qquad \qquad + | \nabla_t \psi|^2 + (W(t) - \epsilon(\beta_0)) |\psi|^2 \big)\, \D s \D t \\   \nonumber
&= \|\phi_{\beta_0} \nabla_t u \|^2 + \|\phi_{\beta_0}K_{\beta_0}u \|^2,
\end{align}
% ------------- %
where we have used the identity
% ------------- %
$$
|\nabla_t(\phi_{\beta_0} u) |^2 = \phi_{\beta_0}^2 |\nabla_t u|^2 + \nabla_t \cdot (\phi_{\beta_0} |u|^2 \nabla_t \phi_{\beta_0}) - |u|^2 \phi_{\beta_0} \Laplace_t \phi_{\beta_0},
$$
% ------------- %
along with the facts that the divergence term integrates to zero in view of the compact support of $u$, and that $\phi_{\beta_0}$ is a ground state of $h_{\beta_0}$ with the energy $\epsilon(\beta_0)$, in other words,
% ------------- %
\begin{align*}
& -|u|^2 \phi_{\beta_0} \Laplace_t \phi_{\beta_0} - \beta_0^2 |u|^2 \phi_{\beta_0} \partial_\varphi^2 \phi_{\beta_0} + W \phi_{\beta_0}^2 |u|^2 \\ & \qquad\quad = |u|^2\phi_{\beta_0}(h_{\beta_0}\phi_{\beta_0}) = \epsilon(\beta_0) \phi_{\beta_0}^2 |u|^2,
\end{align*}
% ------------- %
where $p_\varphi^2 = -\partial_\varphi^2$. The identity \eqref{eq:BetaZeroQuadrForm2}, derived here for all $\psi \in C^\infty_0(\R^3)$, extends by density to all $u$ with $\phi_{\beta_0} u \in \mathcal{Q}(\mathbb{q}_{\beta_0})$.

To get an upper bound to $\Sigma(\beta_0)$, we can simply choose $u(s,t) = g_\zeta(s)$ with $g_\zeta(s) = \ee^{-\zeta s^2}$, which gives
% ------------- %
$$
\mathbb{q}_{\beta_0}[\phi_{\beta_0}g_\zeta] - \epsilon(\beta_0) \|\phi_{\beta_0}g_\zeta\|^2 = 0 + \|\phi_{\beta_0}\|^2 \|g_\zeta'\|^2,
$$
% ------------- %
hence
% ------------- %
$$
\frac{\mathbb{q}_{\beta_0}[\phi_{\beta_0}g_\zeta]}{\|\phi_{\beta_0}g_\zeta\|^2} = \epsilon(\beta_0) + \frac{\|g_\zeta'\|^2}{\|g_\zeta\|^2} \xrightarrow{\zeta \to 0^+}\epsilon(\beta_0) \implies \Sigma(\beta_0) \leq \epsilon(\beta_0).
$$
% ------------- %
As for the corresponding lower bound, $\|\phi_{\beta_0} \nabla_t u \|^2 + \|\phi_{\beta_0}K_{\beta_0}u \|^2$ is clearly  non-negative, thus $\mathbb{q}_{\beta_0}[\phi_{\beta_0}u] \geq \epsilon(\beta_0) \|\phi_{\beta_0}u\|^2 $. Furthermore, since these $\psi$ form a core of $\mathbb{q}_{\beta_0}$, this holds on all the $\mathcal{Q}(\mathbb{q}_{\beta_0})$, giving \mbox{$\Sigma(\beta_0) \geq \epsilon(\beta_0)$.}
\end{proof}

Finally, we will need a result analogous to \cite[Lemma~2]{EKK08}. The latter is a bound stated for an interval $\Lambda\subset\R$; applied at a fixed $t$ and integrated, the same inequality holds in $L^2(\Lambda
\times \R^2, w(t) \D s \D t)$ for any weight in the $t$-variable.
% ------------- %
\begin{lemma}
\label{lemma:technicalLemma}
Let $\Lambda$ be a bounded open interval of $\R$. Then for any open subinterval $\Lambda' \subset \Lambda$ and any $f\in H^1(\Lambda)$ one has
% ------------- %
$$
\|f\|^2_{L^2(\Lambda)} \leq c(\Lambda, \Lambda') \left(\|f\|^2_{L^2(\Lambda')} + \|f'\|^2_{L^2(\Lambda)} \right),
$$
% ------------- %
where $c(\Lambda, \Lambda') \defeq \max \lbrace 2+ 16\left( \frac{|\Lambda|}{|\Lambda'|}\right)^2, 4 |\Lambda|^2 \rbrace$.
\end{lemma}

%%%%%%%%%%%%%%%%%%%%%%%%%%%%%%%%%%%%%%%%%%%%%%%%%%%
\section{Proof of Theorem \ref{theo:TwoParamHardy}}
\label{sec:ProofTwoPar}
\setcounter{equation}{0}

We use the density of $C^\infty_0(\R^3)$ in $H^1(\R^3)$; in view of claim (a) of Lemma \ref{lemma:GroundStateFiber}, it suffices to establish the inequality for $\psi = \phi_0 u$ with $u \in C^\infty_0(\R^3)$. Throughout the discussion we work with the straightened form $\mathbb{q}$ given by \eqref{straightform} and without loss of generality we put $s_0=0$. We fix $b$ with $\mbox{supp } \theta' \subset (-b,b)$ and set $J_s \defeq (-b,b)$, so that the interval in assumption $\mathbf{(T1)}$ satisfies $J_0 \subset J_s$.

The argument proceeds as follows: we use a weighted one-dimensional Hardy inequality to reduce the claim to two estimates, longitudinal derivative (with constant $C_1$) and mass term (with constant $C_2$), both constructed using Lemma \ref{lemma:mixed} below.

\subsection*{One-dimensional Hardy inequality}
The form we need is a slight modification of the conventional inequality \cite[eq.~(26)]{EKK08},
% ------------- %
\begin{equation}
\label{eq:Hardy1dinS}
\int_\R \frac{|v(s)|^2}{1+s^2} \D s \leq 4 \int_\R |v'(s)|^2 \D s, \quad
v \in C^\infty_0(\R), \quad v(0) = 0.
\end{equation}
% ------------- %
To begin with, we put $h(s)=\min \lbrace |s|/b, 1 \rbrace$ and decompose an arbitrary $\psi \in C_0^\infty(\R^3)$ as $ \psi = h \psi + (1-h)\psi$ which allows us to write
% ------------- %
\begin{align*}
\int_{\R\times \R^2} \frac{|\psi(s,t)|^2}{(1+s^2)(1+|t|^2)}\,\D s \D t \leq &\: 2
\int_{\R \times \R^2} \frac{|h\psi|^2}{(1+s^2)(1+|t|^2)}\,\D s \D t \\
&+ 2 \int_{J_s \times \R^2} \frac{|(1-h)\psi|^2}{(1+s^2)(1+|t|^2)}\,\D s \D t.
\end{align*}
% ------------- %
Estimating the first term using \eqref{eq:Hardy1dinS} we get
% ------------- %
\begin{align*}
& 2 \int_{\R^3} \frac{|h\psi|^2}{(1+s^2)(1+|t|^2)}\,\D s \D t \leq \: 8
\int_{\R^3} |\partial_s(h \psi)|^2w(t) \,\D s \D t \\
& \leq 16 \int_{\R^3} |h \partial_s \psi|^2 w(t)\,\D s \D t + 16 \int_{\R^3} |\psi h'|^2w(t)\,\D s \D t,
\end{align*}
% ------------- %
where $w_t = (1+|t|^2)^{-1}$ is the transversal weight. The second term is estimated simply as
% ------------- %
\begin{align*}
\int_{J_s \times \R^2} \frac{|(1-h)\psi|^2}{(1+s^2)} w(t)\,\D s \D t & \leq \int_{J_s \times \R^2} |(1-h)\psi|^2 w(t)\,\D s \D t \\
& \leq  \int_{J_s \times \R^2} |\psi|^2 w(t)\,\D s \D t,
\end{align*}
% ------------- %
where we have used the fact that $|h|\leq 1$. Putting these inequalities together and using $|h'|\le b^{-1}$, we finally arrive at
% ------------- %
\begin{equation}
\label{eq:Final1dHardyinS}
\int_{\R\times \R^2} \frac{|\psi|^2}{1+s^2}\, w(t)\,\D s \D t \leq 16 \|\partial_s
\psi\|^2_{L^2(\R^3,w_t)} + \left(\frac{16}{b^2} + 2 \right) \|\mathbb{1}_{J_s}
\psi \|^2_{L^2(\R^3,w_t)}.
\end{equation}
% ------------- %
To get the result, we have to control both terms on the right hand side by the shifted quadratic form with appropriate constants which we denote as $C_1$ and $C_2$. For convenience we introduce the following notation,
% ------------- %
\begin{align}
I_t &\defeq \|\phi_0 \nabla_t u \|^2, \nonumber \\ \nonumber
I_s &\defeq \|\partial_s \psi \|^2, \\[-.75em] \label{Isymbols} \\[-.75em] \nonumber
I_\varphi &\defeq \| \theta' \partial_\varphi \psi \|^2 , \\ \nonumber
|I_{s\varphi}| &\defeq |2(\phi_0 \partial_s u, \theta' \partial_\varphi
\psi ) |,
\end{align}
% ------------- %
adding the superscript $w$ whenever the weighted version is invoked.

To proceed we need a lemma providing a bound on the mixed term, a necessary ingredient needed to derive both the $C_1$ and $C_2$.
% ------------- %
\begin{lemma}
\label{lemma:mixed}
To any $\alpha,\beta >0$ there exists a number $\gamma_{\alpha,\beta}$ depending
on $\theta_{\mathrm{min}}$,$\|\theta'\|_{L^\infty(J_{s})}$,
$\|\theta''\|_{L^\infty(J_{s})}$, $\lambda$, $c(J_s,J_0)$ and $W$ such that
% ------------- %
$$
|I_{s\varphi}^w| \leq \gamma_{\alpha,\beta} I_t + \alpha I_s^w + \beta
I_\varphi^w.
$$
% ------------- %
\end{lemma}
% ------------- %
The proof is postponed to Section \ref{sec:ProofMixed}; now we show how Lemma \ref{lemma:mixed} yields the said two constants.

\subsection*{Bound on the longitudinal derivative term}
\label{par:BndOnSderivative}
Let us denote $K\defeq (\partial_s - \theta' \partial_\varphi)$ and consider $\psi = \phi_0 u$ with $u \in C^\infty_0(\mathbb{R} \times \mathbb{R}^2)$. For a fixed $s$, we use the identity
% ------------- %
$$
|\nabla_t(\phi_0 u)|^2 = |\phi_0 \nabla_t u|^2 + \nabla_t \cdot (\phi_0 u^2
\nabla_t \phi_0) - u^2 \phi_0 \Delta_t \phi_0.
$$
% ------------- %
Integrating it over $\mathbb{R}^2$, the divergence term vanishes by Gauss's theorem in view of the compact support of $u^2 \phi_0 \nabla_t \phi_0$, while $-u^2 \phi_0\Delta_t \phi_0$ cancels in the quadratic form with $(W - \epsilon_0)|\phi_0 u|^2$ because $H_t \phi_0 = \epsilon_0 \phi_0$; this yields
% ------------- %
$$
\int_{\mathbb{R}^2} \left( |\nabla_t \psi(s,\cdot)|^2 + (W -
\epsilon_0)|\psi(s,\cdot)|^2
\right) \mathrm{d}t = \int_{\mathbb{R}^2} |\phi_0|^2 |\nabla_t u(s,\cdot)|^2 \, \mathrm{d}t =: I_t(s).
$$
% ------------- %
By Fubini theorem, integration over $s$ in combination with \eqref{straightform} then gives $q[\psi] - \epsilon_0 \|\psi\|^2 = I_t + \|K\psi\|^2$.

We focus first on the contribution from the set $J_s \subset \R$. With an abuse of notation we use for its components the `full' symbols \eqref{Isymbols} obtaining
% ------------- %
$$ %\begin{equation}
%\label{eq:mixedFirstIneqStep}
\mathbb{q}[\psi] - \epsilon_0 \|\psi \|^2 = I_t + \| K \psi \|^2 \geq I_t +
\|K \psi \|^2_{w_t} \geq I_t + I_s^w + I_\varphi^w - |I_{s\varphi}^w|.
$$ %\end{equation}
% ------------- %
Using now Lemma \ref{lemma:mixed} with $\alpha<1$ and $\beta=1$, we get
% ------------- %
$$
|I_{s\varphi}^w | \leq \gamma_{\alpha,1} I_t + \alpha I_s^w + I_\varphi^w.
$$
% ------------- %
Writing $I_{s\varphi}^w = \gamma_{\alpha,1}^{-1} I_{s\varphi}^w + (1-\gamma_{\alpha,1}^{-1}) I_{s\varphi}^w $, substituting the above bound for $I_{s\varphi}^w$ in combination with the fact that $\|K \psi \|^2_{w_t} \geq 0$, we have
% ------------- %
\begin{align}
\label{eq:C1partJs}
\mathbb{q}[\psi] - \epsilon_0 \|\psi \|^2
&\geq  I_t + \gamma^{-1} \left( I_s^w + I_\varphi^w - |I_{s\varphi}^w| \right) \\ \nonumber
&\geq I_t + \gamma^{-1}\left( I_s^w + I_\varphi^w - \gamma I_t - \alpha
I_s^w - I_\varphi^w \right) \\ \nonumber
& \geq \gamma^{-1}(1-\alpha) I_s^w,			
\end{align}
% ------------- %
where $\gamma \defeq \max \lbrace \gamma_{\alpha,1},1 \rbrace$. With the same abuse of notation, on $\R \setminus J_s$ we can use the simple bound
% ------------- %
$$
I_s^w \leq \mathbb{q}[\psi] - \epsilon_0 \|\psi \|^2,
$$
% ------------- %
hence together we end up with $C_1 = \frac{\gamma}{1-\alpha}+ 1$.

\subsection*{Bound on the mass term}
A similar strategy can be used to find the constant $C_2$. To this aim, we divide $\|K \psi\|^2$ into two terms and write
% ------------- %
\begin{equation}
\label{eq:C2start}
\mathbb{q}[\psi] - \epsilon_0 \|\psi\|^2 = I_t
+ \frac{1}{2\gamma'} \|K \psi\|^2 + \left(1- \frac{1}{2\gamma'}\right)
\|K \psi\|^2,
\end{equation}
% ------------- %
with $\gamma'$ defined a little below. Using Lemma \ref{lemma:mixed} with $\alpha=1$ and $\beta < 1$, we have
% ------------- %
$$
|I_{s\varphi}^w| \leq \gamma_{1,\beta} I_t + I_s^w + \beta I_\varphi^w.
$$
% ------------- %
We want $\gamma'$ in the above inequality in such a way that $\left(1-\frac{1}{2\gamma'}\right) \geq 0$ as well as $\gamma' \geq \gamma_{1,\beta}$ holds, which is ensured by picking $\gamma' \defeq \max \lbrace \frac12 , \gamma_{1,\beta} \rbrace$. This implies $\left(1-\frac{1}{2\gamma'}\right)\|K\psi ||^2 \geq 0$ and
% ------------- %
\begin{equation}
\label{eq:C2HelperLemma}
|I_{s\varphi}^w| \leq \gamma' I_t + I_s^w + \beta I_\varphi^w.
\end{equation}
% ------------- %
Returning to \eqref{eq:C2start}, we have
% ------------- %
\begin{align*} %\label{mass_term_est}
\mathbb{q}[\psi] - \epsilon_0 \|\psi\|^2 &= I_t
+ \frac{1}{2\gamma'} \|K \psi\|^2 + \left(1- \frac{1}{2\gamma'}\right)
\|K \psi\|^2 \\ \nonumber
&\geq I_t + \frac{1}{2\gamma'} \|K \psi\|^2  \\ \nonumber
&\geq I_t + \frac{1}{2\gamma'} \|K \psi\|^2_{w_t}  \\ \nonumber
&\geq I_t + \frac{1}{2\gamma'} \left(I_s^w + I_\varphi^w - |I_{s\varphi}^w| \right),
\end{align*}
% ------------- %
where we have used successively the inequalities $\left(1-\frac{1}{2\gamma'}\right)\|K\psi ||^2 \geq 0$, $w_t \leq 1$ and $\|K\psi\|^2_{w_t} \geq I_s^w + I_\varphi^w - |I_{s\varphi}^w|$. Furthermore, \eqref{eq:C2HelperLemma} implies
% ------------- %
\begin{align}
\label{eq:C2preLambda}
\mathbb{q}[\psi] - \epsilon_0 \|\psi\|^2  &\geq I_t + \frac{1}{2\gamma'} \left( I_s^w + I_\varphi^w - \gamma' I_t
- I_s^w - \beta I_\varphi^w \right) \\ \nonumber
&= \frac12 I_t + \frac{1-\beta}{2\gamma'} I_\varphi^w,
\end{align}
% ------------- %
where in the last step we utilized Lemma \ref{lemma:lambdaPositive} with $\psi \mapsto \theta'  \psi$, i.e.
% ------------- %
\begin{equation}
\label{eq:C2lambda}
\lambda \|\mathbb{1}_{J_0}\theta' \psi \|^2_{w_t} \leq \|\theta' \upharpoonright J_0 \|^2_\infty I_t + I_\varphi^w.
\end{equation}
% ------------- %
To match the right-hand sides of the inequalities \eqref{eq:C2preLambda} and
\eqref{eq:C2lambda}, we define $k_0\defeq \frac12 \min \lbrace \|\theta'
\upharpoonright J_0 \|^{-2}_\infty, \frac{1-\beta}{\gamma'} \rbrace$, which
allows us to write
% ------------- %
\begin{equation}
\label{eq:C2partJ0}
\mathbb{q}[\psi] - \epsilon_0 \|\psi\|^2  \geq k_0 \left( \|\theta' \upharpoonright J_0\|^2_\infty I_t  + I_\varphi^w\right) \geq \theta^2_{\mathrm{min}} k_0 \lambda \|\mathbb{1}_{J_0} \psi\|^2_{w_t}.
\end{equation}
% ------------- %
The final step is to extend the estimate from $J_0$ to $J_s$ using Lemma \ref{lemma:technicalLemma} with $\Lambda' = J_0$ and $\Lambda= J_s$ and noting that the addition of the weight changes nothing; this yields
% ------------- %
\begin{equation}
\label{eq:C2technical}
\| \mathbb{1}_{J_s}  \psi \|^2_{w_t} \leq c(J_s,J_0) \left( \|\mathbb{1}_{J_0}\psi \|^2_{w_t} + \|\mathbb{1}_{J_s} \partial_s \psi\|^2_{w_t} \right),
\end{equation}
% ------------- %
where the right-hand side is bounded from above by $C_2 \left(\mathbb{q}[\psi] - \epsilon_0 \|\psi \|^2 \right)$, with $C_2 \defeq c(J_s,J_0)\left( \left(\theta^{2}_{\mathrm{min}}k_0 \lambda\right)^{-1} + \frac{\gamma}{1-\alpha} \right)$, using \eqref{eq:C2partJ0} and \eqref{eq:C1partJs}.

\subsection*{The final constant}
Inserting the obtained upper bounds constants with $C_1$ and $C_2$ into \eqref{eq:Final1dHardyinS}, we get
% ------------- %
$$ %\begin{equation}
\int_{\R \times \R^2} \frac{|\psi|^2}{1+s^2} w(t) \D s \D t \leq
\tilde{c}_{\mathrm{H}}^{-1} \left(\mathbb{q}[\psi] - \epsilon_0 \|\psi\|^2 \right),
$$ %\end{equation}
% ------------- %
where
% ------------- %
$$
\tilde{c}^{-1}_{\mathrm{H}} = \left[16 C_1 + \left(\frac{16}{b^2}+2 \right)C_2\right].
$$
% ------------- %

%%%%%%%%%%%%%%%%%%%%%%%%%%%%%%%%%%%%%%%%%%
\section{Proof of Lemma \ref{lemma:mixed}}
\label{sec:ProofMixed}
\setcounter{equation}{0}

For the sake of brevity, we omit the indicator function $\mathbb{1}_{J_{s}}$ further in this proof unless explicitly needed. First of all, we recall that
% ------------- %
\begin{equation}
\label{eq:NablaTBoundAgain}
\|\phi_0 \partial_\varphi u \|^2_{w_t} \leq \| \sqrt{w_t}|t| \phi_0 \nabla_t u
\|^2 \leq  \|\phi_0 \nabla_t u\|^2,
\end{equation}
% ------------- %
since $\frac{|t|^2}{1+|t|^2} \leq 1$. The sought bound concerns the term
% ------------- %
$$ %\begin{equation}
|I_{s\varphi}^w| = |2(\phi_0 \partial_s u, \theta'(\phi_0
\partial_\varphi u + u \partial_\varphi \phi_0))|.
$$ %\end{equation}
% ------------- %
To bound the first summand, we apply inequality \eqref{eq:NablaTBoundAgain} together with Cauchy-Schwarz followed by Young's inequality with $\alpha>0$,
% ------------- %
\begin{align*}
|2( \phi_0 \partial_s u, \theta' \phi_0 \partial_\varphi u)|_{w_t}
&\leq 2 \| \phi_0 \partial_s u \|_{w_t} \|\theta' \phi_0 \partial_\varphi u \|_{w_t}  \\ \nonumber
&\leq 2 \|\theta'\|_\infty\|\phi_0 \nabla_t u \|\|\phi_0 \partial_s u\|_{w_t} \\ \nonumber
&\leq \frac{2c_1}{\alpha} \|\phi_0 \nabla_t u \|^2 + \frac{\alpha}{2} \| \phi_0 \partial_s
u \|_{w_t}^2,
\end{align*}
% ------------- %
where $c_1 \defeq \|\theta'\|^2_\infty $. The bound to the second summand is more involved. We are going to show that $|2(\phi_0 \partial_s u, \theta'u \partial_\varphi \phi_0)| \leq T_2$ with
% ------------- %
$$
T_2 = \tilde{\gamma}_{\alpha,\beta}I_t + \frac{\alpha}{2}I_s^w + \beta I_\varphi^w,
$$
% ------------- %
so that
% ------------- %
$$
|I_{s,\varphi}^w|\leq \frac{2c_1}{\alpha} I_t+ \frac{\alpha}{2} I_s^w + T_2 = \gamma_{\alpha,\beta} I_t + \alpha I_s^w + \beta I_\varphi^w,
$$
% ------------- %
with $\gamma_{\alpha,\beta} = \frac{2c_1}{\alpha}  + \tilde{\gamma}_{\alpha,\beta}$. It remains to determine the $\tilde{\gamma}_{\alpha,\beta}$.

Using twice integration by parts, first with respect to $s$, then to $\varphi$, and noting that both the boundary terms vanish due to the compactness of $u$ in the variable $s$ and the periodicity with respect to $\varphi$, we arrive at the identity
% ------------- %
\begin{align*}
|2(\phi_0 \partial_s u, \theta' u \partial_\varphi \phi_0)| &= |(\phi_0 u,
\theta'' u \partial_\varphi \phi_0)| \\ \nonumber
&= |(\phi_0 u, \theta'' \phi_0 \partial_\varphi u)|,
\end{align*}
% ------------- %
which remains unchanged if one includes the weight $w_t$ since the latter commutes with both the $\partial_\varphi$ and $\partial_s$. By virtue of \eqref{eq:NablaTBoundAgain} and Cauchy-Schwarz again, we get
% ------------- %
$$
|(\phi_0 u, \theta'' \phi_0 \partial_\varphi u)|^2_{w_t} \leq
c_2 \|\phi_0 \nabla_t u\|^2 \|\phi_0 u\|^2_{w_t},
$$
% ------------- %
where $c_2 \defeq 4 \|\theta''\|^2_\infty $.

To find a bound on $\|\mathbb{1}_{J_s} \phi_0 u\|^2_{w_t}$ we employ Lemmata~\ref{lemma:lambdaPositive} and \ref{lemma:technicalLemma}. By assumption $\mathrm{(\mathbf{T1})}$ we have the interval $J_0\subset J_{s}$ such that $\theta_{\mathrm{min}}:=\min_{s\in J_0}|\theta'(s)|>0$. Applying Lemma~\ref{lemma:technicalLemma} with $\Lambda=J_s$ and $\Lambda'=J_0$ and with the weight as in \eqref{eq:C2technical}, we get
% ------------- %
$$
\|\mathbb{1}_{J_s}\phi_0 u\|^2_{w_t}
\le c(J_s,J_0)\Big( \|\mathbb{1}_{J_s}\phi_0 \partial_s u\|^2_{w_t}
+ \theta_{\mathrm{min}}^{-2}\|\mathbb{1}_{J_0}\theta'\phi_0 u\|^2_{w_t}\Big).
$$
% ------------- %
Furthermore, since $\theta'(s) \phi_0(\cdot) u(s,\cdot)  \in H^1(\R^2)$ holds for each fixed $s\in \R$, we can use Lemma \ref{lemma:lambdaPositive} to get
% ------------- %
$$
\|\mathbb{1}_{J_0} \theta' \phi_0 u \|^2_{w_t} \leq
\lambda^{-1}\left(\|\mathbb{1}_{J_{s}}\theta'\partial_\varphi(\phi_0 u)\|^2_{w_t}
+ \|\theta' \|^2_\infty \|\mathbb{1}_{J_{s}} \phi_0 \nabla_t u\|^2\right).
$$
% ------------- %
Combining these estimates, we arrive at
% ------------- %
$$
|2 \left(\phi_0 \partial_s u, \theta' u \partial_\varphi \phi_0 \right)_{w_t}|^2 \leq c_3 \|\theta'\|^2_\infty I_t^2 + c_3 I_t I_\varphi^w + c_3 \lambda \theta^2_{\mathrm{min}}I_t I_s^w,
$$
% ------------- %
with $c_3 \defeq \frac{c_2 c(J_s,J_0)}{\lambda \theta_{\mathrm{min}}^2}$. On the other hand, we have a lower bound on the square of the targeted quantity,
% ------------- %
\begin{align*}
\left(\tilde{\gamma}_{\alpha,\beta} I_t + \frac{\alpha}{2} I_s^w + \beta I_\varphi^w \right)^2
\geq \tilde{\gamma}_{\alpha,\beta}^2 I_t^2 + 2 \tilde{\gamma}_{\alpha,\beta}\beta I_t I_\varphi^w +  \alpha \tilde{\gamma}_{\alpha,\beta} I_t I_s^w
\end{align*}
% ------------- %
where we have dropped the non-negative terms $ \frac{\alpha^2}{4} (I_s^w)^2, \beta^2 (I_\varphi^w)^2$ and $\alpha \beta I_s^w I_\varphi^w$. Comparing the two expressions, we get
% ------------- %
$$ %\begin{equation}
\tilde{\gamma}_{\alpha,\beta}^2 \geq c_3 \|\theta'\|^2_\infty \mbox{, }\alpha \tilde{\gamma}_{\alpha,\beta} \geq c_3 \lambda \theta^2_{\mathrm{min}} \mbox{, } 2\tilde{\gamma}_{\alpha,\beta}\beta \geq c_3.
$$ %\end{equation}
% ------------- %
Requiring all the three inequalities to hold we thus choose
% ------------- %
$$
\tilde{\gamma}_{\alpha,\beta} \defeq \max \Big\lbrace
\sqrt{c_3}\|\theta'\|_\infty, c_3/2\beta, \frac{c_3 \lambda\theta_{\mathrm{min}}^2}{\alpha}
\Big\rbrace.
$$
% ------------- %
With this choice, $|2(\phi_0 \partial_s u, \theta' u \partial_\varphi \phi_0)_{w_t}|^2 \le T_2^2$,
and since both sides are non-negative, taking square roots yields
$|2(\phi_0 \partial_s u, \theta' u \partial_\varphi \phi_0)_{w_t}| \le T_2$.

%%%%%%%%%%%%%%%%%%%%%%%%%%%%%%%%%%%%%%%%%%%%%%%
\section{Proof of Theorem \ref{theo:MainHardy}}
\label{sec:ProofThm2}
\setcounter{equation}{0}

\noindent This proof relies on a combination of Theorem \ref{theo:TwoParamHardy} and the lower bound
% ------------- %
\begin{equation}
\label{eq:CoerciveWeight}
\mathbb{q}[\psi] - \epsilon_0 \|\psi\|^2 \geq \int_\R w_s
h[\psi(s,\cdot)] \D s,
\end{equation}
% ------------- %
which comes from Lemma \ref{lemma:GroundStateFiber} with $\omega= w_s = (1+s^2)^{-1}$. The last piece we need is a lower bound on $H_t-\epsilon_0 +\tilde{c}_{\mathrm{H}} w_t$ in the form sense.
% ------------- %
\begin{lemma}
\label{lemma:pwBoundTransverse}
Let $m_0 \defeq \int_{\R^2} w_t \phi_0^2 \D t \in (0,1]$. Then for every $\nu>0$ we have, in the form sense on $\mathcal{Q}(H_t)$,
% ------------- %
$$
H_t - \epsilon_0 +\nu w_t \geq \Gamma(\nu) =\min \Big\lbrace
\frac{\nu m_0}{4},\frac{\eta m_0}{2(4+m_0)} \Big\rbrace > 0.
$$
% ------------- %
\end{lemma}
% ------------- %
\begin{proof}
Take a normalized vector $u \in \mathcal{Q} \left(H_t\right)$, $\|u\|=1$, of the form $u = a\phi_0+ v$, $a = (\phi_0,u)$, $v \perp \phi_0$, so $|a|^2 + \|v\|^2 =1$, with $v \in \mathcal{Q}\left(H_t\right)$ as well. Substituting into the shifted transverse form, we have
% ------------- %
$$
h[u] = |a|^2 h[\phi_0] + 2 \mbox{Re}\, \overline{a} h(\phi_0, v) + h[v] = h[v] \geq
\eta \|v\|^2,
$$
% ------------- %
where $h(\phi_0,v) = ((H_t - \epsilon_0)\phi_0,v) = 0$, since $H_t \phi_0 = \epsilon_0 \phi_0$ by assumption. We have $0 < w_t \leq 1$ so that $\|w_t^{1/2} v\| \leq \|v\|$, and $\|w_t^{1/2} \phi_0\|^2 = m_0$, hence using triangle inequality we get
% ------------- %
$$
\| w_t^{1/2} u \| \geq (|a|\|w_t^{1/2} \phi_0\| - \|v\|)_+= (|a|\sqrt{m_0} -
\|v\|)_+.
$$
% ------------- %
Note that for any $p,q \geq0$, we have $(p-q)_+^2 \geq p^2 - 2pq$, and thus
% ------------- %
\begin{align*}
\|w_t^{1/2}u\|^2 & \geq m_0 |a|^2 - 2\sqrt{m_0} |a| \|v\| \\ & \geq \frac{m_0}{2}
|a|^2 - 2\|v\|^2 = \frac{m_0}{2} - \frac{4+m_0}{2} \|v\|^2,
\end{align*}
% ------------- %
where we have also used $2pq \leq \frac12 p^2 + 2 q^2$ and $|a|^2 +\|v\|^2 = 1$. Next we set $\delta \defeq \frac{m_0}{2(4+m_0)}$. If $\|v\|^2 \leq \delta$,
then $ \|w_t^{1/2}u\|^2 \geq \frac{m_0}{4}$, and therefore
% ------------- %
$$
h[u] + \nu \|w_t^{1/2}u\|^2 \geq \frac{\nu m_0}{4}.
$$
% ------------- %
On the other hand, if $\|v\|^2 > \delta$, we have
% ------------- %
$$
h[u] + \nu \|w_t^{1/2}u\|^2 \geq \eta \delta = \frac{\eta m_0}{2(4+m_0)},
$$
% ------------- %
hence $h[u] + \nu \|w_t^{1/2} u\|^2 \geq \Gamma(\nu)$ holds in both cases.
\end{proof}

Now we can average two inequalities we have obtained, namely that of Theorem
\ref{theo:TwoParamHardy}, written in the `fiber' form, and
\eqref{eq:CoerciveWeight} to obtain
% ------------- %
$$
\mathbb{q}[\psi] - \epsilon_0 \|\psi\|^2 \geq \frac12 \int_\R w_s (\psi(s,\cdot),
(H_t - \epsilon_0 + \tilde{c}_{\mathrm{H}} w_t) \psi(s,\cdot))_{L^2(\R^2)} \D s.
$$
% ------------- %
For almost every $s$ the fiber $\psi(s,\cdot)$ lies in $\mathcal{Q}(H_t)$ by Lemma~\ref{lemma:GroundStateFiber}, thus by Lemma~\ref{lemma:pwBoundTransverse} with $\nu = \tilde{c}_{\mathrm{H}}$ we get
% ------------- %
$$
h[\psi(s,\cdot)] + \tilde{c}_{\mathrm{H}} (\psi(s,\cdot), w \psi(s,\cdot)) \geq \Gamma(\tilde{c}_{\mathrm{H}})
\|\psi(s,\cdot) \|^2_{L^2(\R^2)} \;\;\mbox{for a.e.}\; s.
$$
% ------------- %
Combining these estimates and integrating over $s$ we get
% ------------- %
$$ %\begin{equation}
\mathbb{q}[\psi] - \epsilon_0 \|\psi\|^2 \geq \frac12 \Gamma(\tilde{c}_{\mathrm{H}})
\int_\R\frac{\|\psi(s,\cdot)\|^2}{1+s^2} \D s = c_{\mathrm{H}}\int_{\R^3}
\frac{|\psi|^2}{1+s^2} \D s \D t,
$$ %\end{equation}
% ------------- %
which is the claim of Theorem~\ref{theo:MainHardy}; Corollary~\ref{cor:WeightCombined} follows by averaging the two theorems.

% ------------- %
\begin{remark}
\label{rem:divprofile}
\rm{The proofs of Theorems~\ref{theo:TwoParamHardy} and \ref{theo:MainHardy} use
	the decay of $W$ only for the identification of the form domain; they
	hold verbatim, with the same constants, for confining profiles with
	$W(t)\to+\infty$ as $|t|\to\infty$ satisfying the regularity requirement
	of $\mathbf{(A_{\mathrm{sa}})}$, with $H^1(\R^3)$ replaced by
	$Q(\mathbb{q})=Q(-\Laplace)\cap Q(W_+)$ in their statements. Indeed,
	writing $W=W_+-W_-$, we have $W_-$ is compactly supported, and thus
	$W_-\in\mathcal{K}_2$, and assuming in addition that $W_+\in
	L^2_{\rm{loc}}(\R^2)$, so that $W_+\in\mathcal{K}_2^{\rm{loc}}$, then
	\cite[Thms.~A.2.7, A.2.8]{Si82} still yield the self-adjointness and
	$C_0^\infty(\R^3)$ as a form core; no growth restriction on $W_+$ is
	required. The existence of $\phi_0$ is then satisfied automatically
	\cite[Thm.~XIII.47]{RS}, as are its simplicity and the positivity gap of
	the $\eta$, the spectrum of $H_t$ being discrete. In this situation it
	is the non-radiality of $W$ only which remains an assumption. Confining transverse profiles of this type belong to standard models of constrained quantum systems, cf.~\cite{WT14}.
}
\end{remark}

%%%%%%%%%%%%%%%%%%%%%%%%%%%%%%%%%%%%%%%%%%%%%%%%%%
\section{Proof of Theorem \ref{theo:SlowDownMean}}
\label{sec:PoTSlowDown}
\setcounter{equation}{0}

Let $H_{\beta_0}$ and $H_\beta$ be as in Section \ref{sec:Main}, and set
$\Sigma(\beta_0) \defeq \inf \sigma(H_{\beta_0})$. By Lemma \ref{lemma:T2IdentifySpectra},  $\Sigma(\beta_0)<0$ and it represents the ground-state energy of the fiber operator $h_{\beta_0}\defeq -\Laplace_t + W+ \beta_0^2 p_\varphi^2$ on $L^2(\R^2)$.
% ------------- %
\begin{remark}
\label{rem:BHappensDetail}
\rm{Before coming to the proof, let us verify that assumption $\mathbf{(B)}$
	holds under any of the two conditions mentioned in
	Remark~\ref{rem:BHappensList}. The first claim follows by testing
	$\mathbb{h}_{\beta_0}[\cdot]$, the quadratic form associated
	with $h_{\beta_0}$, with $\phi_0$. Indeed,
% ------------- %
\begin{align*}
h_{\beta_0}[\phi_0] &= \int_{\R^2} \left(|\nabla_t \phi_0|^2  + W|\phi_0|^2 + \beta_0^2 |p_\varphi \phi_0|^2 \right) \D t \\
&= \int_{\R^2}\left(\epsilon_0 |\phi_0|^2 + \beta_0^2 |p_\varphi \phi_0|^2 \right) \D t\,;
\end{align*}
% ------------- %
requiring this to be negative means $\beta_0^2\|p_\varphi \phi_0\|^2 < |\epsilon_0| \|\phi_0\|^2$. Note that $\epsilon_0<0$ and $\|p_\varphi \phi_0 \|$ is finite in view of the inequality $|\partial_\varphi \phi_0 | \leq |t||\nabla_t\phi_0|$ valid pointwise and the exponential decay of $\phi_0$.

The second claim comes from the observation that if $u\in H^1(\R^2)$ is radial, $ p_\varphi u =0$, and therefore $h_{\beta_0}[u] = (u,H_tu)$. If, in addition, this value is negative, we have
% ------------- %
$$
\inf \sigma(h_{\beta_0})\leq \frac{h_{\beta_0}[u]}{\|u\|^2}= \frac{(u,H_tu)}{\|u\|^2}  <0.
$$
% ------------- %
To show that the stated conditions on the potential $W$ suffice, we test it with $\chi_R$ defined for $R>1$ by
% ------------- %
$$
\chi_R(t) \defeq
\begin{cases}
1, & |t| \leq 1, \\[2pt]
\dfrac{\log (R/|t|)}{\log R}, & 1 < |t| < R, \\[4pt]
0, & |t| \geq R,
\end{cases}
$$
% ------------- %
which is radial and belongs to $H^1(\R^2)$. Since $|\nabla_t \chi_R| = \left(|t|\log R\right)^{-1}$ for $1<|t|<R$ and zero elsewhere, we get
% ------------- %
$$
\|\nabla_t \chi_R\|^2 = \frac{2\pi}{(\log R)^2} \int_{1}^R \frac{\D r}{r} = \frac{2\pi}{\log R} \xrightarrow{R \to \infty} 0.
$$
% ------------- %
Write $W= W_+ - W_-$. Since $\chi_R\to 1$ pointwise as $R\to\infty$ and $W_{\pm} \geq0$, the monotone convergence theorem gives
% ------------- %
$$
\int_{\R^2}W_\pm |\chi_R|^2 \D t \xrightarrow{R \to \infty} \int_{\R^2} W_{\pm} \D t.
$$
% ------------- %
Requiring $W_+ \in L^1(\R^2)$ makes the integral $\int_{\R^2} (W_+ - W_-) \D t$ well defined, thus $\int_{\R^2} W |\chi_R|^2 \D t \xrightarrow{R \to \infty} \int_{\R^2} W \D t$. Putting then the kinetic and potential terms together, we arrive at
% ------------- %
$$
(\chi_R, H_t \chi_R) = \|\nabla_t \chi_R\|^2 + \int_{\R^2} W |\chi_R|^2 \D t \xrightarrow{R \to \infty} 0 + \int_{\R^2}W \D t.
$$
% ------------- %
Thus $(\chi_R, H_t \chi_R)<0$ holds for all $R$ large enough when $\int_{\R^2}W \D t <0$, in particular, for any (nontrivial and) non-positive $W$.
}
\end{remark}
% ------------- %

Regarding the essential spectrum of $H_\beta$, we have the following lemma.
% ------------- %
\begin{lemma}
\label{lemma:MeanEssential}
Under $\mathbf{(A_{\mathrm{sa}})}$, $\mathbf{(B)}$ and $\mathbf{(T2)}$, we have \mbox{$\inf \sigma_{\mathrm{ess}}(H_\beta) \geq \Sigma(\beta_0)$.}
\end{lemma}
% ------------- %
\begin{proof}
By Persson's characterization of the infimum, cf.~\cite[Thm.~3.12]{CFKS87} or \cite{Pe60}, we have
% ------------- %
\begin{align*}
   \inf \sigma_{\mathrm{ess}}(H_\beta) = \lim_{R\to\infty} \inf\Big\lbrace
    \mathbb{q}_\beta [\psi] : \psi \in C^\infty_0(\R^3 \setminus K_R),
    \|\psi \|=1
\Big\rbrace,\\ \nonumber  K_R \defeq [-2R,2R] \times \overline{B_R},
\end{align*}
% ------------- %
where we choose $\mbox{supp}\delta \subset
(-R,R)$ with  $\delta$ being the perturbation from assumption $\mathbf{(T2)}$; we work in the straightened frame, which preserves $K_R$ and the norms. Next we take a fixed $\chi(\cdot) \in C^\infty(\R)$ which is zero on $(-\infty, 1]$ and one on $[2,\infty)$, choose $\xi_1 \defeq \sin \left(\frac{\pi}{2} \chi(|s|/R) \right)$, and set $\xi_2 \defeq \sqrt{1-\xi_1^2}$; obviously $|\xi_1|^2 + |\xi_2|^2 =1$, so it is a valid localization decomposition. With the chosen $\chi$, we have
% ------------- %
$$
|\xi_1'| = \Big| \cos\left(\frac\pi2 \chi(|s|/R) \right) \frac\pi2 \frac1R \chi'\Big| \leq \frac1R \frac\pi2 \|\chi'\|_\infty,
$$
% ------------- %
and therefore
% ------------- %
$$
|\xi_1'|^2 + |\xi_2'|^2 \leq \frac{C}{R^2},
$$
% ------------- %
where $C \defeq\frac14 \pi^2 \|\chi'\|^2_\infty$. For $\psi \in C^\infty_0(K^{\mathrm{c}}_R)$, we then have
% ------------- %
$$
\mbox{supp }(\xi_1 \psi) \subset \lbrace s: |s| > R \rbrace \times \R^2, \;\; \mbox{supp
}(\xi_2 \psi) \subset\lbrace s: |s|< 2R \rbrace \times  \lbrace t: |t|> R\rbrace.	
$$
% ------------- %
because $\psi$ vanishes on $\lbrace (s,t):\, |s| \leq 2R, |t| \leq R\rbrace$. Since the localizers $\xi_1$, $\xi_2$ are independent of $t$, one can commute them with $\nabla_t$ and $\partial_\varphi$, so
% ------------- %
\begin{align*}
\mathbb{q}_\beta [\psi] & = \mathbb{q}_\beta [\xi_1 \psi] + \mathbb{q}_\beta [\xi_2 \psi] - \int_{\R^3} \left(|\xi'_1|^2 + |\xi'_2|^2 \right) |\psi|^2\, \D s\D t \\
& \geq  \mathbb{q}_\beta [\xi_1 \psi] + \mathbb{q}_\beta [\xi_2 \psi] - \frac{C}{R^2} \|\psi\|^2.
\end{align*}
% ------------- %
For $|s| > R$ one has $\theta'_p = \beta_0$, and therefore
% ------------- %
$$
\mathbb{q}_\beta[\xi_1 \psi] \geq \Sigma(\beta_0) \|\xi_1 \psi\|^2.
$$
% ------------- %
On the other hand, for $|t| > R$ we discard the nonnegative kinetic terms and get
% ------------- %
$$
\mathbb{q}_\beta[\xi_2 \psi] \geq \int W |\xi_2 \psi|^2 \geq -\epsilon_R \|\xi_2 \psi\|^2,
$$
% ------------- %
where  $\epsilon_R \defeq \esssup_{|t|>R}W_- \to 0$ as $R \to \infty$ by the assumption on the decay of $W$. Since $\Sigma(\beta_0) < 0 $ by Corollary~\ref{lemma:T2IdentifySpectra}, we have $-\epsilon_R \geq \Sigma(\beta_0)$ for all $R$
large enough. Putting these observations together, we obtain
% ------------- %
$$
\mathbb{q}_\beta[\psi] \geq \Sigma(\beta_0) \|\psi \|^2 - \frac{C}{R^2} \|\psi\|^2,
$$
% ------------- %
and letting $R \to \infty$ we arrive at the sought result.
\end{proof}
% ------------- %

To prove Theorem~\ref{theo:SlowDownMean}, it is by Corollary~\ref{lemma:T2IdentifySpectra} and Lemma~\ref{lemma:MeanEssential} sufficient to find a $\psi \in \mathcal{Q}(\mathbb{q}_\beta)$ with $\mathbb{q}_\beta[\psi] - \Sigma(\beta_0) \|\psi\|^2 <0$. This would imply $\inf \sigma(H_\beta) < \Sigma(\beta_0) \leq \inf \sigma_{\mathrm{ess}} (H_\beta)$, and therefore $\sigma_{\mathrm{disc}}(H_\beta) \neq \emptyset$. We define the appropriate trial function as
% ------------- %
\begin{equation}
\psi(s,t) \defeq g(s) \phi_{\beta_0}(t),
\end{equation}
% ------------- %
where $g \in H^1(\mathbb{R})$ is a longitudinal profile to be determined, noting that $g\phi_{\beta_0} \in \mathcal{Q}(\mathbb{q}_\beta)$, and $\|p_\varphi \phi_{\beta_0}\|> 0$ as in Lemma~\ref{lemma:PhiNoRotSymm}.
For this $\psi$ we evaluate the quadratic form $\mathbb{q}_\beta[\psi]$ in the straightened frame,
% ------------- %
$$
\mathbb{q}_\beta[\psi] = \int_{\mathbb{R}^3} \left( |(\partial_s - \theta'(s) \partial_\varphi) \psi|^2 + |\nabla_t \psi|^2 + W(t)|\psi|^2 \right) \D s \D t.
$$
% ------------- %
Using $p_\varphi = -i\partial_\varphi$ and the ansatz $\psi = g \phi_{\beta_0}$, we have $(\partial_s - \theta' \partial_\varphi) (g \phi_{\beta_0})= g'(s) \phi_{\beta_0}(t) - i \theta'(s) g(s) (p_\varphi \phi_{\beta_0})(t)$, so the first expression in the above integral can be rewritten as
% ------------- %
$$
|(\partial_s - \theta' \partial_\varphi) \psi|^2 = |g'|^2 |\phi_{\beta_0}|^2 + \theta'(s)^2 |g|^2 |p_\varphi \phi_{\beta_0}|^2 - 2 \mathrm{Re}\left( i \overline{g'} g \theta' \overline{\phi_{\beta_0}} (p_\varphi \phi_{\beta_0}) \right).
$$
% ------------- %
The cross-term vanishes since
% ------------- %
$$
\int_{\mathbb{R}^2} \phi_{\beta_0} (p_\varphi \phi_{\beta_0}) \D t = -i
\int_{\mathbb{R}^2} \phi_{\beta_0} (\partial_\varphi \phi_{\beta_0}) \D t =
-\frac{i}{2}\int_{\mathbb{R}^2}\, \partial_\varphi (\phi_{\beta_0}^2) \D t = 0
$$
% ------------- %
and the form reduces to
% ------------- %
\begin{align*}
\mathbb{q}_{\beta}[\psi] &= \int_{\mathbb{R}} |g'(s)|^2 \D s + \int_{\mathbb{R}} \theta'(s)^2 |g(s)|^2 \|p_\varphi \phi_{\beta_0}\|^2_{L^2_t} \D s \nonumber \\
&\quad + \int_{\mathbb{R}} |g(s)|^2 \left( \int_{\mathbb{R}^2} \left( |\nabla_t \phi_{\beta_0}|^2 + W |\phi_{\beta_0}|^2 \right) \D t \right) \D s.
\end{align*}
% ------------- %
By the definition of the state $\phi_{\beta_0}$ we have
% ------------- %
$$
\left( \phi_{\beta_0}, H_t \phi_{\beta_0} \right) = \Sigma(\beta_0)  - \beta_0^2 \|p_\varphi \phi_{\beta_0}\|^2.
$$
% ------------- %
and substituting this back into the expression for $\mathbb{q}_\beta[\psi]$, we obtain
% ------------- %
\begin{align*}
\mathbb{q}_{\beta}[\psi] &= \int_{\mathbb{R}} |g'|^2 \D s + \int_{\mathbb{R}} |g|^2 \left( \theta'(s)^2 \|p_\varphi \phi_{\beta_0}\|^2 + \Sigma(\beta_0) - \beta_0^2 \|p_\varphi \phi_{\beta_0}\|^2 \right) \D s \nonumber \\
&= \Sigma(\beta_0) \|g\|^2_{L^2(\mathbb{R})} + \int_{\mathbb{R}} |g'(s)|^2 \D s + \|p_\varphi \phi_{\beta_0}\|^2 \int_{\mathbb{R}} (\theta'(s)^2 - \beta_0^2) |g(s)|^2 \D s.
\end{align*}
% ------------- %
Subtracting next the threshold energy $\Sigma(\beta_0) \|\psi\|^2 = \Sigma(\beta_0) \|g\|^2$, we find
% ------------- %
$$ %\begin{equation}
\mathbb{q}_\beta[\psi] - \Sigma(\beta_0) \|\psi\|^2 = \int_{\mathbb{R}} |g'(s)|^2 \D s + \int_{\mathbb{R}} V_{\mathrm{eff}}(s) |g(s)|^2 \D s,
$$ %\end{equation}
% ------------- %
where $V_{\mathrm{eff}}(s) \defeq C_0 (\theta'(s)^2 - \beta_0^2)$ with the
constant $C_0 \defeq \|p_\varphi \phi_{\beta_0}\|^2 > 0$.

This reduces the problem to the existence of a negative-energy state for the one-dimensional Schr\"odinger operator $-\frac{\D^2}{\D s^2} + V_{\mathrm{eff}}(s)$. By assumption, $\int_{\mathbb{R}} (\theta'(s)^2 - \beta_0^2) \D s < 0$, which implies $\int_{\mathbb{R}} V_{\mathrm{eff}}(s)\, \D s < 0$. Explicitly, choosing the family of scaled Gaussians, $g_\zeta(s) = \ee^{-\zeta s^2}$, the kinetic energy term vanishes as $\mathcal{O}(\sqrt{\zeta})$, while
the potential energy term converges to $\int V_{\mathrm{eff}} \D s < 0$ by dominated convergence, since $V_{\mathrm{eff}} \in L^1(\R)$ by $\mathbf{(T2)}$. This ensures that the form is negative for all sufficiently small $\zeta$, hence $\inf \sigma (H_\beta) < \Sigma(\beta_0) \leq \inf \sigma_{\mathrm{ess}}(H_\beta)$ holds by Lemma~\ref{lemma:MeanEssential}, and we conclude that $\sigma_{\mathrm{disc}}(H_\beta) \neq \emptyset$.

%%%%%%%%%%%%%%%%%%%%%%%%%%%%%%%%%%%%%%%%%%%%%%%%%%%%%%%%%%%%%%%%

\section*{Acknowledgment}
\noindent The authors are indebted to Vladimir Lotoreichik for useful discussions.

\end{document}